\documentclass{JHEP3} 

\usepackage{mcite}

\usepackage{amsmath, amssymb}
\usepackage{concmath, palatino}
\usepackage{mathrsfs}

\usepackage{graphicx,epsfig}
\usepackage{epic}
\usepackage{color}

\newcommand\fverb{\setbox\pippobox=\hbox\bgroup\verb}
\newcommand\fverbdo{\egroup\medskip\noindent%
			\fbox{\unhbox\pippobox}\ }
\newcommand\fverbit{\egroup\item[\fbox{\unhbox\pippobox}]}
\newbox\pippobox

\newcommand{\be}{\begin{equation}}
\newcommand{\ee}{\end{equation}}
\newcommand{\ba}{\begin{eqnarray}}
\newcommand{\ea}{\end{eqnarray}}

\newcommand{\la}{\longrightarrow}

\newcommand{\ads}{$AdS_5\times S^5$\ }

\allowdisplaybreaks

    \newcommand{\beq}{\begin{equation}}
    \newcommand{\eeq}{\end{equation}}
    \newcommand\beqa{\begin{eqnarray}}
    \newcommand\eeqa{\end{eqnarray}}

\newcommand{\CS}{\mathcal{S}}
\newcommand{\CJ}{\mathcal{J}}

\title{Quantum folded string in $S^{5}$ and the Konishi multiplet at strong coupling}

\author{Matteo Beccaria\\
  Dipartimento di Fisica, Universita' del Salento \& INFN, \\
  Via Arnesano, 73100 Lecce, Italy\\
  E-mail: \email{matteo.beccaria$\bullet$le.infn.it}
}

\author{Guido Macorini\\
Niels Bohr Institute (NBI), University of Copenhagen, \\
Blegdamsvej 17, DK-2100 Copenhagen, Denmark\\
E-mail: \email{guido.macorini$\bullet$le.infn.it}
}

\abstract{
The Konishi superconformal multiplet is an important theoretical laboratory where 
one can test AdS/CFT methods to compute strong coupling corrections to the spectrum of 
superstrings in \ads. In particular, one can exploit integrability for finite charge states/operators. The multiplet ground state is a singlet operator with two simple descendants in the rank-1 sectors $\mathfrak{sl}(2)$ and $\mathfrak{su}(2)$
of $\mathcal N=4$ super Yang-Mills theory. Recently, the next-to-leading quantum correction to the $\mathfrak{sl}(2)$
state has been computed. Here, we use the algebraic curve approach to determine the correction to the 
other state recovering universality of the correction inside the multiplet.
}

\begin{document} 

\section{Introduction and result}
\label{sec:intro}

AdS/CFT correspondence~\cite{Maldacena:1997re, *Witten:1998qj, *Gubser:1998bc} relates the spectrum of 
conformal dimensions of the ${\cal N}=4$ SYM theory to the spectrum of $AdS_5 \times S^5$ superstring.
In the planar limit integrability emerges \cite{Lipatov:1993yb, *Faddeev:1994zg, *Minahan:2002ve, *Beisert:2003tq, *Bena:2003wd, *Kazakov:2004qf} and anomalous dimensions can be computed as eigenvalues of an integrable 
super spin chain by solving nested non-perturbative Bethe Ansatz equations 
\cite{Beisert:2003yb, *Beisert:2005fw, *Beisert:2006ez}. These equations are asymptotic, {\em i.e.} valid for states with large enough charges.
Finite charge states are more difficult and their anomalous dimensions, including the so-called 
wrapping corrections, are captured by the Y-system \cite{Gromov:2009tv, *Bombardelli:2009ns, *Gromov:2009bc, *Arutyunov:2009ur, Cavaglia:2010nm} successfully checked at strong coupling in the quasi-classical
limit  \cite{Gromov:2009tq, *Gromov:2010vb}. At weak-coupling the leading order 
predictions from the Y-system \cite{Gromov:2009tv} agree with standard field theoretical calculations
 \cite{Fiamberti:2007rj, *Velizhanin:2008jd}. At next-to-leading order they are also in agreement
 \cite{Arutyunov:2010gb, *Balog:2010xa, *Balog:2010vf}  with the L\"uscher corrections  \cite{Bajnok:2008bm, *Bajnok:2009vm}.

\medskip
Beyond perturbation theory the Y-system can be treated numerically. The anomalous dimension of the states in the
Konishi multiplet have been an important theoretical laboratory to test the method.
In \cite{Gromov:2009tv} the Y-system was combined with the vacuum TBA equations to produce an infinite set of integral equations for the $\mathfrak{sl}(2)$ sector of the
spectrum. They  were then solved numerically for the simplest state in the Konishi multiplet
 \cite{Gromov:2009zb}. The numerical approach starts in the weak-coupling regime and pushes the 
  `t Hooft coupling $\lambda$ to large values in order to extrapolate to the strong-coupling limit
  \cite{Gromov:2009zb, Frolov:2010wt}. The prediction obtained in
\cite{Gromov:2009zb} for the Konishi anomalous dimension $\gamma$ is
\be
\label{eq:konishi} 
\gamma+4=2.0004\lambda^{1/4}+1.99/\lambda^{1/4}+\cdots\,\, .
\ee 
The leading coefficient agrees with the prediction of
\cite{Gubser:2002tv} giving $2$.  This was also confirmed in a recent paper \cite{Passerini:2010xc} in the light-cone approach.
  
\medskip
The problem with an analytical proof of a relation like (\ref{eq:konishi}) is only technical, but very hard.
In particular, it is expected that the  analytical structure of the Y-system at finite coupling 
\cite{Cavaglia:2010nm} becomes very complicated at strong coupling. 

\medskip
A very interesting approach, pioneered by A. Tseytlin and collaborators, 
is based on the semiclassical quantization
of spinning string solutions with large charges and recently systematically applied to the 
problem of the Konishi multiplet in \cite{Roiban:2009aa, Roiban:2011fe}.
To explain the basic idea we can consider the simple case of the spinning folded string  with  two charges, 
the Lorentz spin $S$ and R-charge $J$. Let us introduce the ratios
$\CS=S/\sqrt{\lambda},\ \CJ=J/\sqrt{\lambda}$.  
If we expand at large $\lambda$ and fixed $\CS, \CJ$, the expansion of
the energy is of the form 
\be
\label{enrgyexpnt} 
E\equiv
\gamma+S+J=\sqrt{\lambda}\,E_0(\CS,\CJ)+E_1(\CS,\CJ)+\frac
1{\sqrt{\lambda}}\,E_{2}(\CS,\CJ)+\ldots\ .  
\ee 
If we now replace the ratios $\CS, \CJ$ by their definitions, fix $S$ and $J$, and re-expand at large $\lambda$
we find  that the above
expansion turns into a  power series of the type \cite{Roiban:2009aa}
\be
\label{eq:largelambda}
E=\lambda^{1/4}a_0+\frac{1}{\lambda^{1/4}}a_2+\cdots\,.
\ee
Here, the classical energy $E_0$ contributes to the first coefficient $a_0$ while
both $E_1$, the one-loop $\sigma$-model correction,  and $E_0$ contribute to the coefficient $a_2$. 
Eq.~(\ref{eq:largelambda}) is indeed the expected near-flat space large $\lambda$
expansion for the energy of a finite charge state. Although the above result is obtained from a semiclassical calculation
where $S, J$ are always large, it is tempting to identify $a_{0}$, $a_{2}$
with the coefficients of the expansion of the finite charge state. 

\medskip
The advantage of this approach is that all calculations can be done by semiclassical methods in the string theory or, 
exploiting the integrability structures, by working with the simpler quasi-classical Y-system whose equivalence with 
the semiclassical computation has been established in \cite{Gromov:2009tq}.
The short string expansion of the energy for the $(S, J)$ folded string
reads (see for instance  \cite{Roiban:2009aa, Roiban:2011fe})
\be
\label{eq:short-sl2}
E^{\mathfrak{sl}(2)}(S,  J) = \sqrt{2\,S}\,\lambda^{1/4}\,\left[
1+\frac{1}{\sqrt\lambda}\left(
\underbrace{\frac{3\,S}{8}+\frac{J^{2}}{4\,S}}_{\rm classical}+
\underbrace{a_{01}^{\mathfrak{sl}(2)}}_{\rm quantum}
\right)
\right]+\dots\, .
\ee
In this expression, the terms labeled {\em classical} come from the expansion of the classical energy. The one-loop corrections are fully encoded in the {\em quantum} term $a_{01}^{\mathfrak{sl}(2)}$.
The algebraic curve quantization procedure for an arbitrary
$\CS$ and $\CJ$ \cite{SchaferNameki:2010jy,Gromov:2008ec} leads to the result \cite{Gromov:2011de}
\be
a_{01}^{\mathfrak{sl}(2)} = -\frac{1}{4}.
\ee 
The Konishi state is associated with $S=J=2$ and we obtain an analytical prediction for 
the coefficients in Eq.~(\ref{eq:largelambda}) in full agreement with the numerical results of \cite{Gromov:2009zb} (see also \cite{Roiban:2011fe,Vallilo:2011fj}), 
\be
\label{eq:universal}
a_{0} = a_{2} = 2.
\ee

\medskip
It is very interesting to study the manifestation of superconformal invariance at the level of  strong coupling
corrections. The multiplet structure can 
be regarded as a consistency check of any method attempting to deal with such regime. This problem has been addressed
in \cite{Roiban:2009aa, Roiban:2011fe} from the perspective of semiclassical string quantization. Here, we would like to test the algebraic curve approach from this point of view. To this aim, we remind that quantum string states as well as dual gauge theory operators are highest weight states with Dynkin labels
\be
[p_{1}, q, p_{2}]_{\left(s_{L}, s_{R}\right)},
\ee
where, in terms of the classical charges $S_{1,2}, J_{1, 2, 3}$ , 
the $\mathfrak{so}(4)=\mathfrak{su}(2)\oplus \mathfrak{su}(2)$ labels $(s_{L}, s_{R})$ 
are given by $s_{L, R}=\frac{1}{2}(S_{1}\pm S_{2})$ and the Dynkin 
labels $[p_{1}, q, p_{2}]$ of $\mathfrak{su}(4)$ are given by  $p_{1,2} = J_{2}\mp J_{3}$, $q=J_{1}-J_{2}$. 

\noindent
With this notation, the singlet operator $\mbox{Tr}(\overline\Phi^{i}\,\Phi_{i})$ with bare dimension 2 
is the top state $[0,0,0]_{(0,0)}$
of the Konishi multiplet. It has two superconformal descendants in the $\mathfrak{sl}(2)$ and $\mathfrak{su}(2)$
sectors given by the following  states with bare dimension 4
\be
\begin{array}{ccc}
{\rm sector} & {\rm state} & [p_{1}, q, p_{2}]_{s_{L}, s_{R}} \\
\hline
\mathfrak{sl}(2) & \mbox{Tr}(\Phi_{1}\,D^{2}\,\Phi_{1}) & [0,2,0]_{(1,1)} \\
\mathfrak{su}(2) & \mbox{Tr}([\Phi_{1}, \Phi_{2}]^{2}) & [2,0,2]_{(0,0)} 
\end{array}
\ee
The  state in the $\mathfrak{sl}(2)$ sector has been worked out in details in \cite{Gromov:2011de}. As is well known
it is associated with a classical string solution represented by a string rotating in just one plane in $S^{5}$ with a 
spin in $AdS_{5}$ \cite{Frolov:2002av}. We shall denote is as the $(S, J)$ folded string.
The second state has been discussed in details in \cite{Beisert:2003xu, Beisert:2003ea} and it is associated with
a classical string rotating in two planes in $S^{5}$, the $(J_{1}, J_{2})$ folded string \cite{Frolov:2003xy}. 
The two (classical) solutions are related by an analytic continuation connecting the respective string profiles and conserved charges.
From the point of view of the Bethe Ansatz description, at least in the gauge theory, they are quite different.
The folded $(S, J)$ string is described by a 2-cut solution with symmetric cuts on the real axis. Instead, 
the folded $(J_{1}, J_{2})$ string is  associated (at least at weak coupling) with a 2-cut solution with two cuts symmetric around the 
imaginary axis and with a non-trivial geometry. The special role of these particular very symmetric 
2-cut solutions has been investigated in details in \cite{Vicedo:2007rp}.

\medskip

\medskip
It is very interesting to pursue the duality in the context of the algebraic curve approach
(or the equivalent quasi-classical Y-system). In particular, one would like to check whether the multiplet structure is obeyed by the first non trivial strong coupling correction to the energy.
A first analysis in this direction has been presented in  \cite{Roiban:2009aa, Roiban:2011fe}. The 
one-loop corrected energy for the $(J_{1}, J_{2})$ folded string takes a form similar to Eq.~(\ref{eq:short-sl2})
\ba
E^{\mathfrak{su}(2)}(J_{1}, J_{2}) &=& \sqrt{2\,J_{2}}\,\lambda^{1/4}\,\left[
1+\frac{1}{\sqrt\lambda}\left(
\underbrace{\frac{J_{2}}{8}+\frac{J_{1}^{2}}{4\,J_{2}}}_{\rm classical}+
\underbrace{a_{01}^{\mathfrak{su}(2)}}_{\rm quantum}
\right)
\right]+\dots\,.
\ea
The authors of \cite{Roiban:2009aa, Roiban:2011fe}
conjectured that $a_{01}^{\mathfrak{su}(2)}$ should be the same with an opposite sign , {\em i.e.} $+\frac{1}{4}$,
reflecting the opposite sign of the curvature of $S^{3}$ as compared to $AdS_{3}$.
This proposal is consistent with similar behaviour of the correction for circular spinning strings \cite{Roiban:2009aa, Roiban:2011fe}.
For the Konishi representative with $J_{1}=J_{2}=2$, the assignment  $a_{01}^{\mathfrak{su}(2)}=\frac{1}{4}$ 
leads to the same strong coupling correction as for the $\mathfrak{sl}(2)$ Konishi descendant
\be
E^{S=2, \,J=2} = E^{J_{1}=2, \,J_{2}=2} = 2\,\lambda^{1/4}+\frac{2}{\lambda^{1/4}}+\cdots.
\ee
Beyond the Konishi state, this choice is also consistent with the superconformal degeneracy of the states 
$(S=2, J)$ and $(J_{1}=J, J_{2}=2)$~\footnote{
It follows for instance by duality of the Bethe equations and adding roots at infinity to implement superconformal transformations.} because it predicts the same correction~\footnote{The case $S=2, J=3$ has been 
confirmed by an independent TBA computation in \cite{Gromov:2011de}.}
\be
E^{S=2, \,J} = E^{J_{1}=J, \,J_{2}=2} = 2\,\lambda^{1/4}+\frac{J^{2}+4}{4}\,\frac{1}{\lambda^{1/4}}+\cdots.
\ee
Finally, as a further support of the conjecture $a_{01}^{\mathfrak{su}(2)}=\frac{1}{4}$, we recall
that an argument in \cite{Roiban:2011fe}~\footnote{We thank A. Tseytlin and R. Roiban for pointing out this issue.}
suggests that the independence of $a_{01}$ on the charge ratio is not accidental and has instead a deep origin
being related to the continuity of observables with respect to the addition of a small  charge to the principal
one ($S$ or $J_{2}$ for the two folded strings). 

\medskip
In this paper we perform an algebraic curve calculation of the correction and provide very convincing 
numerical evidence that the 
result $a_{01}^{\mathfrak{su}(2)}=\frac{1}{4}$ proposed in 
\cite{Roiban:2009aa, Roiban:2011fe} is indeed correct.

\section{Algebraic curve method for the \ads \ superstring}
\label{sec:general-ac}

The general construction of the algebraic curve for the \ads superstring is discussed for instance in 
\cite{Gromov:2007aq,Gromov:2008ec}. Here, we summarize in a self-contained way the main results for the reader's convenience.

\subsection{Classical algebraic curve}

The monodromy matrix of the Lax connection for the integrable dynamics of the \ads superstring has eigenvalues
\be
\{e^{i\,\widehat p_{1} }, e^{i\,\widehat p_{2} }, e^{i\,\widehat p_{3} }, e^{i\,\widehat p_{4} } |
e^{i\,\widetilde p_{1} }, e^{i\,\widetilde p_{2} }, e^{i\,\widetilde p_{3} }, e^{i\,\widetilde p_{4} }\}
\ee
The eigenvalues are roots of the characteristic polynomial and define an 8-sheeted Riemann surface. The classical
algebraic curve has macroscopic cuts connecting various pairs of sheets. Around each cut, we have 
\be
p^{+}_{i}-p^{-}_{j} = 2\,\pi\,n_{ij},\qquad x\in\mathcal C^{ij}_{n},
\ee
where $n$ is an integer associated with the cut. The possible combinations of sheets (a.k.a. polarizations) that are relevant for \ads are
\be
i=\widetilde 1, \widetilde 2, \widehat 1, \widehat 2,\qquad
j=\widetilde 3, \widetilde 4, \widehat 3, \widehat 4.
\ee
The properties of the monodromy matrix implies (for folded configurations) the inversion properties
\ba
\label{eq:inversion}
\widetilde p_{1,2}(x) &=& -2\,\pi\,m-\widetilde p_{2,1}(1/x),\qquad m\in\mathbb Z, \nonumber \\
\widetilde p_{3,4}(x) &=& +2\,\pi\,m-\widetilde p_{4,3}(1/x),\\ 
\widehat p_{1,2,3,4}(x) &=& -\widehat p_{2,1,4,3}(1/x). \nonumber
\ea
The poles of the connection plus Virasoro constraints implies the pole structure around the special points 
$x=\pm 1$~\footnote{At weak coupling, the two points collapse and we end with the usual pole at $x=0$
well known in the study of integrable spin chains.},
\be
\{\widehat p_{1}, \widehat p_{2}, \widehat p_{3}, \widehat p_{4} |
\widetilde p_{1}, \widetilde p_{2}, \widetilde p_{3}, \widetilde p_{4}\} \sim 
\frac{\{\alpha_{\pm}, \alpha_{\pm}, \beta_{\pm}, \beta_{\pm} |\alpha_{\pm}, \alpha_{\pm}, \beta_{\pm}, 
\beta_{\pm}\}}{x\pm 1}.
\ee
Also, the asymptotic value at $x\to \infty$ is related to the conserved charges as in ($\mathcal Q = \frac{Q}{\sqrt \lambda}$)
\be
\left(
\begin{array}{c}
\widehat p_{1} \\
\widehat p_{2} \\
\widehat p_{3} \\
\widehat p_{4} \\
\hline
\widetilde p_{1} \\
\widetilde p_{2} \\ 
\widetilde p_{2} \\ 
\widetilde p_{4}  
\end{array}
\right) = \frac{2\pi}{x}\left(\begin{array}{c}
+\mathcal E-\mathcal S_{1}+\mathcal S_{2} \\
+\mathcal E+\mathcal S_{1}-\mathcal S_{2} \\
-\mathcal E-\mathcal S_{1}-\mathcal S_{2} \\
-\mathcal E+\mathcal S_{1}+\mathcal S_{2} \\
\hline
+\mathcal J_{1}+\mathcal J_{2}-\mathcal J_{3} \\
+\mathcal J_{1}-\mathcal J_{2}+\mathcal J_{3} \\
-\mathcal J_{1}+\mathcal J_{2}+\mathcal J_{3} \\
-\mathcal J_{1}-\mathcal J_{2}-\mathcal J_{3} 
\end{array}\right)+\mathcal O(1/x^{2}),
\ee

\section{Fluctuations frequencies from the algebraic curve}

The macroscopic cuts can be thought as the condensation of a large number of poles as it happens in semiclassical  quantum mechanics for a large excitation number. We shall be interested in the effect of the addition of a single pole and in the shift $p\to p+\delta p$ of the quasi-momenta. This insertion will compute the quantum fluctuations around the classical solution. From the definition of the action-angle variables for the integrable string, we deduce that residue of $\delta p$ around such a pole has to be
\be
\label{eq:pole}
\delta p \sim \pm\frac{\alpha(x_{p})}{x-x_{p}},\qquad \alpha(x) = \frac{4\pi}{\sqrt\lambda}\frac{x^{2}}{x^{2}-1}.
\ee
The position of the poles can be found by solving (for generic $n$) the equation 
\be
p_{i}(x_{n}^{ij})-p_{j}(x^{ij}_{n}) = 2\pi\ n, \qquad |x_{n}^{ij}|>1,
\ee
for all polarizations $(i,j)$ with $i<j$ and the pairs
\ba
S^{5} &:& (i,j) = (\widetilde 1, \widetilde 3), (\widetilde 1, \widetilde 4), (\widetilde 2, \widetilde 3), (\widetilde 2, \widetilde 4), \\
AdS_{5} &:& (i,j) = (\widehat 1, \widehat 3), (\widehat 1, \widehat 4), (\widehat 2, \widehat 3), (\widehat 2, \widehat 4), \\
\mbox{Fermions} &:& (i,j) =  (\widetilde 1, \widehat 3), (\widetilde 1, \widehat 4), (\widetilde 2, \widehat 3), (\widetilde 2, \widehat 4), \\
&& \phantom{(i,j) =}\  (\widehat 1, \widetilde 3), (\widehat 1, \widetilde 4), (\widehat 2, \widetilde 3), (\widehat 2, \widetilde 4).\nonumber
\ea

The correction to the quasi-momenta will be $\delta p_{i}$ with the pole structure (\ref{eq:pole}), regularity across the macroscopic cuts, and asymptotic behaviour ($N_{ij} = \sum_{n} N_{n}^{ij}$ is the number of $(i,j)$ excitations)
\be
\delta \left(
\begin{array}{c}
\widehat p_{1} \\
\widehat p_{2} \\
\widehat p_{3} \\
\widehat p_{4} \\
\hline
\widetilde p_{1} \\
\widetilde p_{2} \\ 
\widetilde p_{3} \\ 
\widetilde p_{4}  
\end{array}
\right) = \frac{4\pi}{x\,\sqrt\lambda}\left(\begin{array}{c}
+\frac{1}{2}\delta\Delta+N_{\widehat 1\, \widehat 4}+N_{\widehat 1\, \widehat 3}+N_{\widehat 1\, \widetilde 3}+N_{\widehat 1\, \widetilde 4} \\
+\frac{1}{2}\delta\Delta+N_{\widehat 2\, \widehat 4}+N_{\widehat 2\, \widehat 3}+N_{\widehat 2\, \widetilde 3}+N_{\widehat 2\, \widetilde 4} \\
-\frac{1}{2}\delta\Delta-N_{\widehat 2\, \widehat 3}-N_{\widehat 1\, \widehat 3}-N_{\widetilde 1\, \widehat 3}-N_{\widetilde 2\, \widehat 3} \\
-\frac{1}{2}\delta\Delta-N_{\widehat 1\, \widehat 4}-N_{\widehat 2\, \widehat 3}-N_{\widetilde 2\, \widehat 4}-N_{\widetilde 1\, \widehat 4} \\
\hline
\ \ \ \ \ \ \ \ \ \ \ \ -N_{\widetilde 1\, \widetilde 4}-N_{\widetilde 1\, \widetilde 3}-N_{\widetilde 1\, \widehat 3}-N_{\widetilde 1\, \widehat 4} \\
\ \ \ \ \ \ \ \ \ \ \ \ -N_{\widetilde 2\, \widetilde 3}-N_{\widetilde 2\, \widetilde 4}-N_{\widetilde 2\, \widehat 4}-N_{\widetilde 2\, \widehat 3} \\
\ \ \ \ \ \ \ \ \ \ \ \ +N_{\widetilde 2\, \widetilde 3}+N_{\widetilde 1\, \widetilde 3}+N_{\widehat 1\, \widetilde 3}+N_{\widehat 2\, \widetilde 3} \\
\ \ \ \ \ \ \ \ \ \ \ \ +N_{\widetilde 1\, \widetilde 4}+N_{\widetilde 2\, \widetilde 4}+N_{\widehat 2\, \widetilde 4}+N_{\widehat 1\, \widetilde 4} 
\end{array}\right)+\mathcal O(1/x^{2}),
\ee
The precise values of the residues can be read off the definition of the action-angle variables and are  
\be
\mathop{\mbox{res}}_{x=x^{ij}_{n}} \widehat p_{k} = (\delta_{i\,\widehat k}-\delta_{j\,\widehat k})\,\alpha(x^{ij}_{n})\,N_{n}^{ij}, \qquad
\mathop{\mbox{res}}_{x=x^{ij}_{n}} \widetilde  p_{k} = (\delta_{i\,\widetilde k}-\delta_{j\,\widetilde k})\,\alpha(x^{ij}_{n})\,N_{n}^{ij}, 
\ee
where $k=1,2,3,4$, and $i<j$ taking values $\widehat 1, \widehat 2, \widehat 3, \widehat 4, \widetilde 1, 
\widetilde 2, \widetilde 3, \widetilde 4$.
The anomalous shift $\delta\Delta$ can be written as a linear combination of the $N^{ij}$ numbers
\be
\delta\Delta = \sum_{n, (ij)} N^{ij}_{n}\,\Omega^{ij}_{n}.
\ee
This formula for $\delta \Delta$ exhibits the classical frequencies $\Omega^{ij}$ around the classical solution. 
These frequencies can be thought as normal mode frequencies. After quantization, and taking into account statistics, 
the one loop correction to the energy can be written as a sum over zero  point energies
\be
\label{eq:oneloop-correction}
\delta E = \frac{1}{2}	\sum_{n, (ij)}(-1)^{F}\,\Omega_{n}^{ij}.
\ee

\subsection{Inversion symmetry and linear combinations of frequencies for rank-1 solutions}

The inversion symmetry (\ref{eq:inversion}) implies the two important
relations
\ba
\Omega^{\widetilde 1\, \widetilde 4}(x) &=& -\Omega^{\widetilde 2\,\widetilde 3}(1/x)+
\Omega^{\widetilde 2\,\widetilde 3}(0),  \\
\Omega^{\widehat 1\, \widehat 4}(x) &=& -\Omega^{\widehat 2\,\widehat 3}(1/x)-2.
\ea
In addition, we have linear relations between the various $\Omega^{ij}$ which can be easily read 
by representing a particular frequency connecting two sheets as the sum of the intermediate frequencies
connecting an intermediate sheet.
Assuming the top-down symmetry (valid for rank-1 solutions)
\be
p_{\widehat 1, \ \widehat 2, \ \widetilde 1, \ \widetilde 2} = -p_{\widehat 4, \ \widehat 3, \ \widetilde 4, \ \widetilde 3},
\ee
one can prove that all the $8+8$ physical frequencies can be written in terms of the two 
basic ones
\be
\Omega_{S}(x) = \Omega^{\widetilde 2\,\widetilde 3}(x),\qquad
\Omega_{A}(x) = \Omega^{\widehat 2\,\widehat 3}(x).
\ee
The final result is 
\ba
&& \Omega^{\widetilde 1\, \widetilde 4} = -\Omega_{S}(1/x)+\Omega_{S}(0),\\
&& \Omega^{\widetilde 2\, \widetilde 4} = \Omega^{\widetilde 1\, \widetilde 3} = 
\frac{1}{2}[\Omega_{S}(x)-\Omega_{S}(1/x)+\Omega_{S}(0)],\\
&& \Omega^{\widehat 1\, \widehat 4} = -\Omega_{A}(1/x)-2, \\
&& \Omega^{\widehat 2\, \widehat 4} = \Omega^{\widehat 1\,\widehat 3} = 
\frac{1}{2}[\Omega_{A}(x)-\Omega_{A}(1/x)]-1,\\
&& \Omega^{\widehat 2\, \widetilde 4} = \Omega^{\widetilde 1\, \widehat 3} = 
\frac{1}{2}[\Omega_{A}(x)-\Omega_{S}(1/x)+\Omega_{S}(0)],\\
&& \Omega^{\widetilde 2\, \widehat 4} = \Omega^{\widehat 1\, \widetilde 3} = 
\frac{1}{2}[\Omega_{S}(x)-\Omega_{A}(1/x)]-1,\\
&& \Omega^{\widetilde 1\, \widehat 4} =\Omega^{\widehat 1\, \widetilde 4} = 
\frac{1}{2}[-\Omega_{S}(1/x)-\Omega_{A}(1/x)+\Omega_{S}(0)]-1, \\
&& \Omega^{\widehat 2\, \widetilde 3} = \Omega^{\widetilde 2\, \widehat 3} = 
\frac{1}{2}[\Omega_{S}(x)+\Omega_{A}(x)].
\ea

\section{Algebraic curve computation for the $(J_{1}, J_{2})$ folded strings}
\label{sec:}

\subsection{Classical $(S, J)$ folded string in the short string limit}

According to \cite{Beisert:2003ea}, the folded string rotating in $AdS_{5}$ and $S^{5}$ with angular momenta $S$ and $J$ can be analitically continued to the folded string
rotating in $S^{5}$ with two angular momenta $J_{1}$, $J_{2}$ according to the replacement rule
\be
\label{eq:continuation}
(E, J_{1}, J_{2}) \leftrightarrow (-J, -E, S).
\ee 
In the $(S, J)$ folded string, the two cuts of the elliptic curve are symmetrically placed along the real axis, 
$(a,b)$, $(-a,-b)$, where $1<a<b$. The conserved quantities are given by the expressions \cite{Gromov:2011de}
\ba
S &=& 2\,n\,g\,\frac{ab+1}{ab}\,\left[b\,\mathbb E\left(1-\frac{a^{2}}{b^{2}}\right)
-a\,\mathbb K\left(1-\frac{a^{2}}{b^{2}}\right)\right], \nonumber \\
J &=& \frac{4\,n\,g}{b}\,\sqrt{(a^{2}-1)(b^{2}-1)}\,\mathbb K\left(1-\frac{a^{2}}{b^{2}}\right), \\
E &=& 2\,n\,g\frac{ab-1}{ab}\,\left[b\,\mathbb E\left(1-\frac{a^{2}}{b^{2}}\right)
+a\,\mathbb K\left(1-\frac{a^{2}}{b^{2}}\right)\right].\nonumber
\ea
The branch points can be expanded for small $S$ and $J$ according to
\ba
a &=& 1+\frac{\rho ^2 s^3}{8}+\frac{1}{128} \left(\rho ^2-\rho ^4\right) s^5+\frac{\rho ^4 s^6}{128}+\frac{\rho ^2 \left(4 \rho ^4-22 \rho ^2-9\right) s^7}{4096}+\mathcal O\left(s^8\right), \\
b &=& 1+2 s+2 s^2+\frac{1}{8} \left(\rho ^2+7\right) s^3+\frac{1}{4} \left(\rho ^2-1\right) s^4+\frac{1}{256} \left(-2 \rho ^4+34 \rho ^2-85\right) s^5+\mathcal O\left(s^6\right).\nonumber
\ea
Indeed, the associated charges are 
\ba
S &=& 2\,n\,\pi\,g\,s^{2}+\mathcal O(s^{6}), \nonumber \\
J &=& 2\,n\,\pi\,g\,\rho\,s^{2}+\mathcal O(s^{7}), \\
E &=&4\,n\,\pi\,g\,s+\frac{1}{4} \pi  g n \left(2 \rho ^2+3\right) s^3-\frac{1}{128} s^5 \left(\pi  g n \left(4 \rho ^4-20 \rho ^2+21\right)\right)+\mathcal O\left(s^6\right).\nonumber
\ea 
This $s\sim \sqrt S$ and $\rho=\frac{J}{S}$. More precisely, from $\sqrt\lambda = 4\,\pi\,g$, we have 
\be
\label{eq:spin-s}
s = \sqrt{\frac{S}{2\,n\,\pi\,g}} = \frac{\sqrt{2S/n}}{\lambda^{1/4}}.
\ee
The short string expansion of the energy is
\be
\frac{E}{n\sqrt \lambda} = s+\frac{1}{16} \left(2 \rho ^2+3\right) s^3+\frac{1}{512} \left(-4 \rho ^4+20 \rho ^2-21\right) s^5+\mathcal O\left(s^6\right).
\ee

\subsection{Analytic continuation to the $(J_{1}, J_{2})$ folded string}

In order to describe the $(J_{1}, J_{2})$ string, we apply the 
 continuation (\ref{eq:continuation}) and are now led to study
\ba
J_{1} &=& -2\,n\,g\,\frac{ab-1}{ab}\,\left[b\,\mathbb E\left(1-\frac{a^{2}}{b^{2}}\right)
+a\,\mathbb K\left(1-\frac{a^{2}}{b^{2}}\right)\right], \nonumber \\
J_{2} &=& 2\,n\,g\,\frac{ab+1}{ab}\,\left[b\,\mathbb E\left(1-\frac{a^{2}}{b^{2}}\right)
-a\,\mathbb K\left(1-\frac{a^{2}}{b^{2}}\right)\right], \\
E &=& -\frac{4\,n\,g}{b}\,\sqrt{(a^{2}-1)(b^{2}-1)}\,\mathbb K\left(1-\frac{a^{2}}{b^{2}}\right).\nonumber
\ea
We expand $a$, $b$ around the point $-1$. Introducing the small parameter $s$, we find the expansion
\ba
a &=&  -1+i s+\left(\frac{1}{2}-\frac{\rho }{2}\right) s^2+\frac{1}{16} i (8 \rho -3) s^3+\frac{1}{16} \left(-2 \rho ^2+2 \rho -1\right) s^4+\nonumber \\
&& + \frac{1}{512} i \left(32 \rho ^2+16 \rho +3\right) s^5+\mathcal O\left(s^6\right) ,\nonumber \\
b &=& \overline a.
\ea
Notice that this is precisely the short string limit of the double contour discussed in \cite{Beisert:2003xu}. These 
branch points give
\ba
\label{eq:exp1}
J_{2} &=& 2\,n\,\pi\,g\,s^{2}+\mathcal O(s^{6}), \nonumber \\
J_{1} &=& 2\,n\,\pi\,g\,\rho\,s^{2}+\mathcal O(s^{7}), \\
E &=& 4 \,n\,\pi\,  g \, s+\frac{1}{4} g n \left(2 \pi  \rho ^2+\pi \right) s^3-\frac{1}{128} s^5 \left(\pi  g n \left(4 \rho ^4-28 \rho ^2-3\right)\right)+\mathcal O\left(s^6\right).\nonumber
\ea
Using again the relation (\ref{eq:spin-s}) and identifying $\rho=\frac{J_{1}}{J_{2}}$,
we find the following expansion
\be
\label{eq:exp2}
\frac{E}{n\sqrt \lambda} = s+\frac{1}{16} \left(2 \rho ^2+1\right) s^3+\frac{1}{512} \left(-4 \rho ^4+28 \rho ^2+3\right) s^5+\mathcal O\left(s^6\right).
\ee
It can be easily shown that this result is in full agreement with the general treatment in \cite{Vicedo:2007rp}.

\subsection{Construction of the $p_{\widetilde 2}$ quasi-momentum}
\label{sec:construction}

In \cite{Gromov:2011de}, the reader can find the explicit non-trivial quasimomentum $p_{\widehat 2}$ for the 
$(S,J)$ folded string. Following our approach based on the analitic continuation, we can look for a suitable 
continuation of it. As we can verify {\em a posteriori} (see the Appendix), this procedure gives the 
sphere quasi-momentum $p_{\widetilde 2}$. The result is 
(written here with the standard branch line assignment for the square root)
\ba
\label{eq:spherep2}
p_{\widetilde 2}&=& \pi\,n-i\,\frac{\Delta}{2\,g}\left(\frac{a}{a^{2}-1}-\frac{x}{x^{2}-1}\right)\,\sqrt{\frac{b}{a}\,\frac{a^{2}-1}{b^{2}-1}}\sqrt{\frac{|a|-i\,a}{|a|-i\,\overline a}\,\frac{\overline a-x}{a-x}}\,
\sqrt{\frac{a}{\overline a}\frac{|a|-i\,a}{|a|-i\,\overline a}\,\frac{\overline a+x}{a+x}}+\nonumber\\
&& -\frac{2 a b J_{2}}{g\,(b-a)(a b+1)}\,F_{1}(x)-\frac{\Delta\,(a-b)}{2\,g\,\sqrt{(a^{2}-1)(b^{2}-1)}}\,F_{2}(x), \\
F_{1}(x) &=& i\,\mathbb F\left(i\,\sinh^{-1}\sqrt{-\frac{a-b}{a+b}\,\frac{a-x}{a+x}}, \frac{(a-b)^{2}}{(a+b)^{2}}
\right), \\
F_{2}(x) &=& i\,\mathbb E\left(i\,\sinh^{-1}\sqrt{-\frac{a-b}{a+b}\,\frac{a-x}{a+x}}, \frac{(a-b)^{2}}{(a+b)^{2}}
\right),
\ea
where
\ba
J_{1} &=& +2\,n\,g\,\frac{ab-1}{ab}\,\left[b\,\mathbb E\left(1-\frac{a^{2}}{b^{2}}\right)
+a\,\mathbb K\left(1-\frac{a^{2}}{b^{2}}\right)\right], \nonumber \\
J_{2} &=& -2\,n\,g\,\frac{ab+1}{ab}\,\left[b\,\mathbb E\left(1-\frac{a^{2}}{b^{2}}\right)
-a\,\mathbb K\left(1-\frac{a^{2}}{b^{2}}\right)\right], \\
\Delta &=& -\frac{4\,n\,g}{b}\,\sqrt{(a^{2}-1)(b^{2}-1)}\,\mathbb K\left(1-\frac{a^{2}}{b^{2}}\right).\nonumber
\ea
This expression is valid provided 
\be
\mbox{Re}(a), \mbox{Im}(a)>0, \qquad b = -\overline a.
\ee
It can be checked that 
\be
p(a) = p(\overline a) = n\,\pi,\qquad
p(-a) = p(-\overline a) = -n\,\pi,\qquad
p(\infty) = 0.
\ee
Setting 
\be
a =  1+i s+\frac{1}{2} (\rho -1) s^2+\frac{1}{16} i (8 \rho -3) s^3+\frac{1}{16} \left(2 \rho ^2-2 \rho +1\right) s^4+\frac{1}{512} i \left(32 \rho ^2+16 \rho +3\right) s^5+O\left(s^6\right),
\ee
we recover the expansion (\ref{eq:exp1}) and (\ref{eq:exp2}). Notice again that in this section $b=-\overline a$
and {\bf not} $b=\overline a$ as in the previous section. This is necessary to have the correct cut structure.

\medskip
\noindent
The full set of quasi-momentum is obtained by completing the sphere quasi-momenta with the 
relations 
\be
p_{\widetilde 2}(x) = -p_{\widetilde 3}(x) = -p_{\widetilde 1}(1/x) = p_{\widetilde 4}(1/x),
\ee
and by assigning the following AdS quasi-momenta (following from the absence of cuts in the AdS sheets)
\be
p_{\widehat 1, \widehat 2} = -p_{\widehat 3, \widehat 4} = 
\frac{\Delta}{2g}\,\frac{x}{x^{2}-1} = \frac{2\,\pi\,\mathcal E\,x}{x^{2}-1}.
\ee

\medskip
\noindent
The  sphere quasi-momentum $p_{\widetilde 2}$ defined in (\ref{eq:spherep2}) has branch cuts along small arcs of circumference
with radius $|a|$. A typical plot of it has the form shown in Fig.~(\ref{fig:p2-arcs}) where we can see the cuts
and the singularity around $x=\pm 1$.
Actually, these are not the physical branch cuts ~\cite{Bargheer:2008kj}.

\FIGURE{
\label{fig:p2-arcs}
\epsfig{width=10cm,file=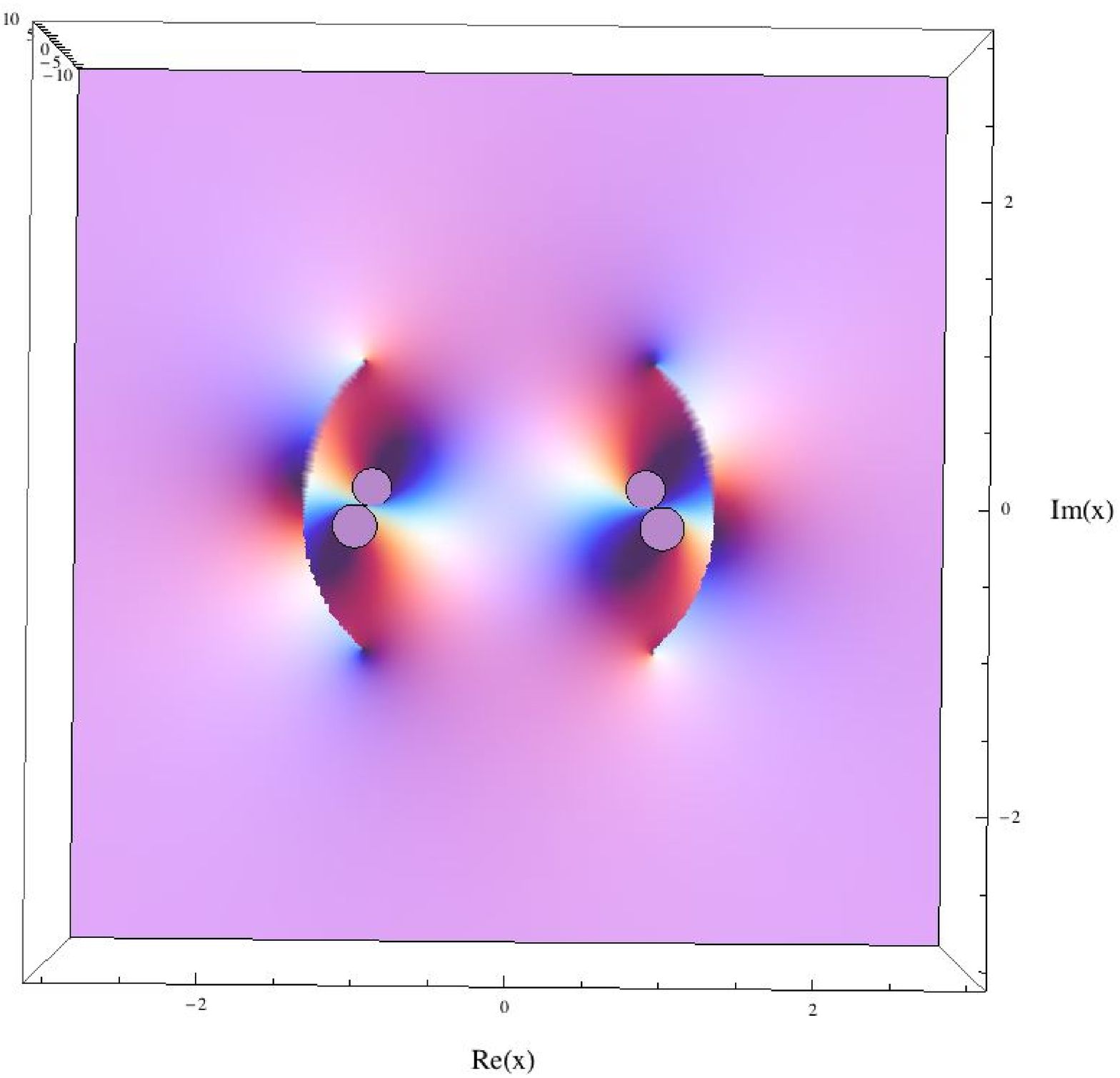}
\caption{Typical form of $p_{\widetilde 2}$. In the plot, we can recognize the cuts along arcs of circumference 
$|x|=|a|$ as well as the poles at $x=\pm 1$.
}
}

\subsection{Fluctuation energies for the $(J_{1}, J_{2})$ folded string}

The general structure of quantum fluctuations around symmetric 2-cut $\mathfrak{su}(2)$ solutions
has been investigated in detail in  \cite{Gromov:2008ec}. 
The fluctuations of quasi momenta with excitation of type 
$(\widehat 2, \widehat 3)$ (with $N^{\widehat 2\,\widehat 3}=1$) at $z$ and 
excitation of type 
$(\widetilde 2, \widetilde 3)$ (with $N^{\widetilde 2\,\widetilde 3}=1$) at $y$
have the general form 
\ba
\delta p_{\widehat 2} &=& \frac{\alpha(z)}{x-z}+\frac{\delta\alpha_{-}}{x-1}+\frac{\delta\alpha_{+}}{x+1}, \\
\delta p_{\widetilde 2} &=& \frac{1}{f(x)}\left[
-\frac{f(y)\,\alpha(y)}{x-y}+\frac{\delta\alpha_{-}\,f(1)}{x-1}+\frac{\delta\alpha_{+}\,f(-1)}{x+1}-\frac{4\pi\,x}{\sqrt\lambda}+A\right],
\ea
where $\delta\alpha_{\pm}$ and  $A$ are constants and $f(x)^{2} = (x-a)(x-\overline a)(x-b)(x-\overline b)$. Using the inversion relations and replacing in the asymptotic condition we easily find 
\ba
\delta\Delta &=& \Omega_{S}(y)+\Omega_{A}(z),\\
\Omega_{A}(x) &=& \frac{2}{x^{2}-1}\left(1+x\,\frac{f(1)-f(-1)}{f(1)+f(-1)}\right), \\
\Omega_{S}(x) &=& \frac{4}{f(1)+f(-1)}\left(\frac{f(x)}{x^{2}-1}-1\right).
\ea
The considered solutions have the additional symmetry 
\be
p_{\widetilde 2} = -p_{\widetilde 3},\qquad
p_{\widetilde 1} = -p_{\widetilde 4},\qquad
p_{\widehat 1} = p_{\widehat 2} = -p_{\widehat 3} = -p_{\widehat 4}.
\ee
This we can identify all frequencies with the above pairing of indices. Consistency requires the following relation which indeed is true for the above expressions
\be
\Omega_{A}(x)+\Omega_{A}(1/x)+2=0.
\ee
We end with the following simple expressions for the six independent frequencies

\medskip
\noindent
{\bf Bosonic fluctuations}
\ba
\Omega_{S} &=& \Omega^{\widetilde 2\, \widetilde 3}, \\
\Omega_{\overline S} &=& \Omega^{\widetilde 1\, \widetilde 4} = -\Omega_{S}(1/x)+\Omega_{S}(0),\\
2\times \Omega_{S_{\perp}} &=& \Omega^{\widetilde 2\, \widetilde 4} = \Omega^{\widetilde 1\, \widetilde 3} = 
\frac{1}{2}[\Omega_{S}(x)-\Omega_{S}(1/x)+\Omega_{S}(0)],\\
4\times \Omega_{A} &= & \Omega^{\widehat 1\, \widehat 4} = 
\Omega^{\widehat 2\, \widehat 4} = \Omega^{\widehat 1\,\widehat 3} = \Omega^{\widehat 2\,\widehat 3}, 
\ea

\medskip
\noindent
{\bf Fermionic fluctuations}
\ba
4\times \Omega_{\overline F}  &=& \Omega^{\widehat 2\, \widetilde 4} = \Omega^{\widetilde 1\, \widehat 3} = 
\Omega^{\widetilde 1\, \widehat 4} =\Omega^{\widehat 1\, \widetilde 4} = 
\frac{1}{2}[\Omega_{A}(x)-\Omega_{S}(1/x)+\Omega_{S}(0)],\\
4\times \Omega_{F} &=& \Omega^{\widetilde 2\, \widehat 4} = \Omega^{\widehat 1\, \widetilde 3} = 
\Omega^{\widehat 2\, \widetilde 3} = \Omega^{\widetilde 2\, \widehat 3} = 
\frac{1}{2}[\Omega_{S}(x)+\Omega_{A}(x)].
\ea

\section{Evaluation of the one-loop correction}
\label{sec:}

The standard way to compute the one-loop energy (\ref{eq:oneloop-correction}) is to write the sum over the mode number $n$ as a contour integral
\be
\delta E = \frac{1}{2}	\sum_{ij} (-1)^{F_{ij}}\oint\frac{dx}{2\,\pi\,i}\left(
\Omega^{ij}(x)\,\partial_{x}\,\log\,\sin\frac{p_{i}-p_{j}}{2}.
\right)
\ee
The integral is conveniently computed by deforming the contour in two pieces:
\begin{enumerate}
\item[a)] The unit circumference $|x|=1$, 
\item[b)] a contour surrounding the cut in the $(\widetilde 2, \widetilde 3)$ plane.
\end{enumerate}

\medskip
The (a) contribution is rather easy. All singularities cancel as a consequence of the ultraviolet finiteness of the 
correction. The (b) contribution is less trivial since it requires some insight about how to deform the contour integration 
around the cut. In the simplest case with mode number $n=1$, the one relevant for 
Konishi~\footnote{Notice that for $n>1$, the structure of fluctuations becomes more complicated. 
In particular, there are $n-1$ additional fluctuations near each branch cut endpoint. We shall not discuss configurations
with $n>1$ here. For these states one has to identify the precise contour integration around the cut.
}, we 
find the structure of excitations for the $(\widetilde 2, \widetilde 3)$ polarization shown in Fig.~(\ref{fig:contour}).
\FIGURE{
\label{fig:contour}
\epsfig{width=10cm,file=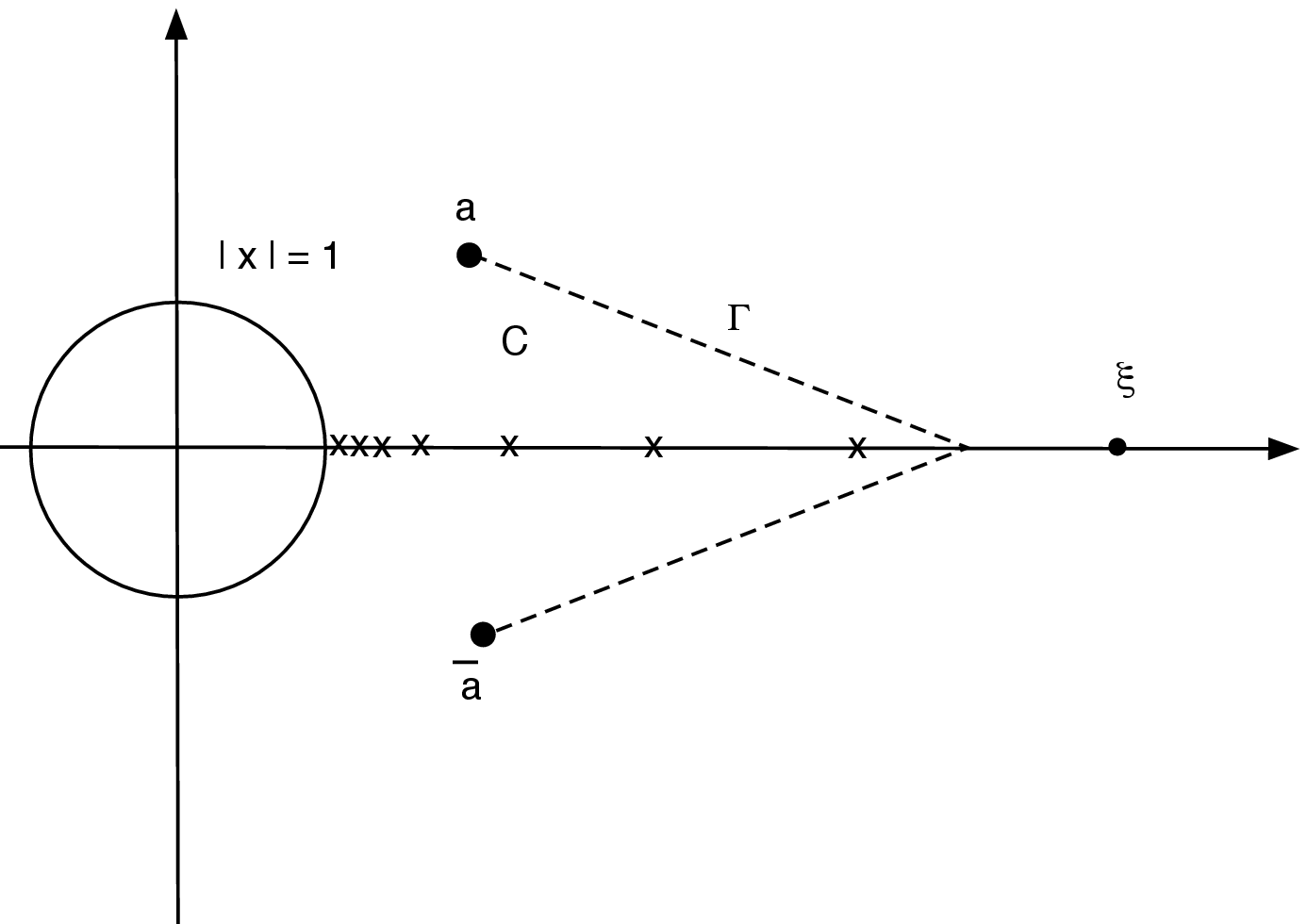}
\caption{Sketch of the fluctuations poles appearing for the $n=1$ folded string. Crosses denote an infinite sequence
accumulating at the point $x=1$. The point $\xi$ is a somewhat separate pole whose position depends on $\rho$
and moves to infinity as $\rho\to 1$. Similar poles can be found for $\mbox{Re} x <0$ and are not drawn.}
}
Apart from the cut endpoints, we only find poles on the real axis. All but one of them can be grouped in an infinite sequence $\{x_{k}\}$ that accumulates at $x=1$ with  $|x_{k}-1|\sim 1/k$ for large $k$.
Then, there is a somewhat different pole at $x=\xi$ whose position
depends on $\rho$ and tends to infinity $\xi\to \infty$ for $\rho\to 1$. A specific example of this structure is shown in Fig.~(\ref{fig:poles}),  where we show the plot of $\mbox{Re}\,\partial_{x}\log\sin p_{\widetilde 2}$ at $s=1/10$ and $\rho=2$.
\FIGURE{
\label{fig:poles}
\epsfig{width=7cm,file=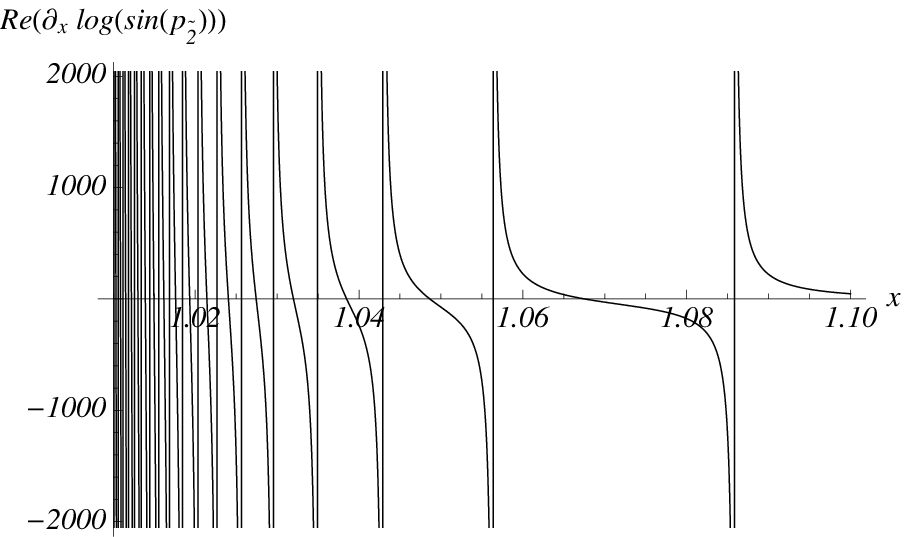}
\epsfig{width=7cm,file=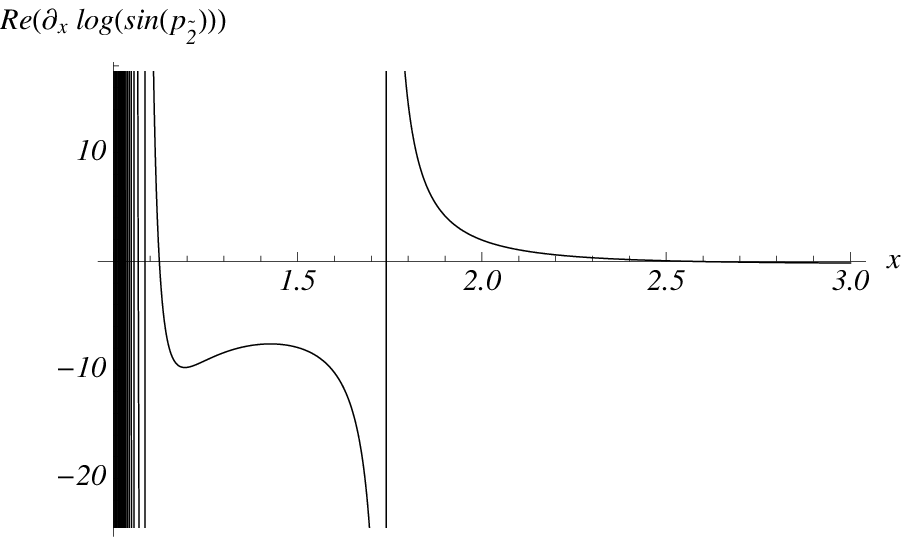}
\caption{$\mbox{Re}\,\Omega_{S}$ for $s=\frac{1}{5}$ and $\rho=1$. The pinches are an infinite sequence of 
poles condensating around $x=1$ plus two poles at $x=a, \overline a$. The various white lines are artifacts of the 
plot.
}
}
The left and right panels differ by the range of $x$. The left plot focuses on the region near $x=1$ and shows 
the regular infinite sequence of poles $\{x_{k}\}$. The right plot shows that there is a pole at $x=\xi=1.74078$
well separated from the other poles. Its contribution is non zero and must be included.



In order to evaluate the cut integral we continue the quasi-momentum the right in the complex plane. 
We deform suitably the integration contour and compute the discontinuity taking into account the jump of sign of $f(x)$ across the physical cut. From the computational point of view it is convenient to compute the integral along the dashed polygonal $\Gamma$ in Fig.~(\ref{fig:contour}) and evaluate separately
the contribution of the special pole $\xi$. This is particularly important for values of $\rho$ near 1, the Konishi case,
when $\xi$ is large.

%

\medskip
We collect in Fig.~(\ref{fig:plots}) a few sample numerical values of the one-loop correction evaluated at the special values of the ratio of the two spins $\rho=1, \frac{3}{2}, 2$ as a function of the spin parameter $s$.
\FIGURE{
\label{fig:plots}
\epsfig{width=12cm,file=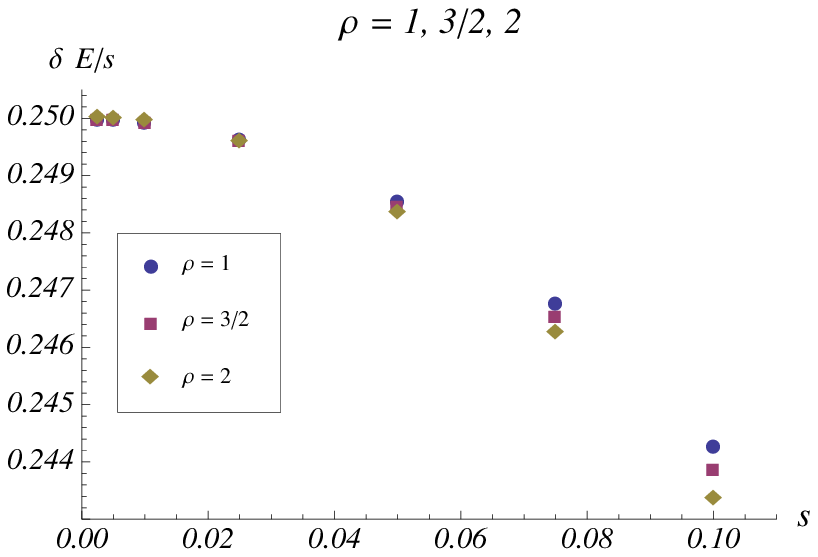}
\caption{Ratio $\delta E/s$ for $\rho=1, 3/2, 2$.}
}
The independence of the $s\to 0$ limit with respect to $\rho$ is clear. 
A simple polynomial fit to a similar larger set of points provides the following estimate of the  $s\to 0$ limit
\be
\lim_{s\to 0}\frac{\delta E}{s} = 0.24999999(1).
\ee
where the error represents the dependence on $\rho$. Our computations clearly provides 
strong evidence for the correctness of the exact result $a_{01}^{\mathfrak{su}(2)}=\frac{1}{4}$.
%

\section{Conclusions}

In this paper we have computed the strong coupling next-to-leading correction to the energy of the
quantum folded string with two angular momenta $J_{1, 2}$ in $S^{5}$ in the limit where 
$J_{2}$ is large with fixed ratio $J_{1}/J_{2}$ and small $J_{1,2}/\sqrt\lambda$ (semiclassical short string limit).
This state is expected to capture at semiclassical level the properties of the  $\mathfrak{su}(2)$ descendent of the Konishi state. It belongs to the same
multiplet as the analogous state dual to the folded string with spin in $AdS$ and angular momentum in $S^{5}$. The correction should be the same as a consequence of superconformal symmetry. 

\medskip
We performed the computation by exploiting the algebraic curve method proposed in \cite{Gromov:2008ec}. We computed the one-loop correction numerically with high precision
and confirmed the conjecture proposed in  \cite{Roiban:2009aa, Roiban:2011fe}.

\medskip
A natural continuation of this work is of course to perform a similar analysis for the small circular strings
solutions considered in \cite{Roiban:2009aa, Roiban:2011fe} in order to prove universality of the next-to-leading 
strong coupling correction for more states with bare dimension 4 in the Konishi multiplet. This analysis is in progress.

\medskip
Finally, the structure of multiplets beyond Konishi seems unclear at
the moment so it is important to collect as much data as possible to
see if there are degeneracies
in energy for various other semiclassical states with different
quantum numbers. The algebraic curve approach is clearly a powerful tool in this respect.

\section*{Acknowledgments}

We thank Nikolay Gromov for many clarifications about the algebraic curve approach to one-loop corrections
and for help in correcting a mistake in the first version of this paper. We thank 
Arkady Tseytlin and Radu Roiban for important comments and for suggesting non-trivial consistency checks.
We also thank   
Fedor Levkovich-Maslyuk for helpful discussions.

\appendix
\section{Asymptotics of $p_{\widetilde 2}$}

Evaluating $p_{\widetilde 2}$ at large $x$ we find
\ba
p_{\widetilde 2}(x) &\stackrel{x\to\infty}{\la}& \frac{J_{2}-J_{1}}{2\,g\,x} = \frac{2\pi}{x}(\mathcal J_{2}-\mathcal J_{1}), \\
p_{\widetilde 2}(1/x) &\stackrel{x\to 0}{\la}& -\frac{J_{2}+J_{1}}{2\,g}\,x .
\ea
Comparing with the general asymptotic behaviour 
\be
\left(
\begin{array}{c}
\widehat p_{1} \\
\widehat p_{2} \\
\widehat p_{3} \\
\widehat p_{4} \\
\hline
\widetilde p_{1} \\
\widetilde p_{2} \\ 
\widetilde p_{2} \\ 
\widetilde p_{4}  
\end{array}
\right) = \frac{2\pi}{x}\left(\begin{array}{c}
+\mathcal E-\mathcal S_{1}+\mathcal S_{2} \\
+\mathcal E+\mathcal S_{1}-\mathcal S_{2} \\
-\mathcal E-\mathcal S_{1}-\mathcal S_{2} \\
-\mathcal E+\mathcal S_{1}+\mathcal S_{2} \\
\hline
+\mathcal J_{1}+\mathcal J_{2}-\mathcal J_{3} \\
+\mathcal J_{1}-\mathcal J_{2}+\mathcal J_{3} \\
-\mathcal J_{1}+\mathcal J_{2}+\mathcal J_{3} \\
-\mathcal J_{1}-\mathcal J_{2}-\mathcal J_{3} 
\end{array}\right)+\mathcal O(1/x^{2}),
\ee
and with the inversion properties
\ba
\widetilde p_{1,2}(x) &=& -\widetilde p_{2,1}(1/x)-2\,\pi\,m, \\
\widetilde p_{3,4}(x) &=& -\widetilde p_{4,3}(1/x)+2\,\pi\,m, \\
\widehat p_{1,2,3,4}(x) &=& -\widehat p_{2,1,4,3}(1/x), 
\ea
we identify the correct asymptotic behaviour of $ \widetilde p_{2}=-\widetilde p_{3}$ after exchanging $J_{1}\leftrightarrow J_{2}$.

\bibliography{AdS-CFT}{}
\bibliographystyle{JHEP}

\end{document}